\documentclass[10pt,showpacs,twocolumn]{revtex4-2}
\usepackage{color}
\usepackage{graphicx}
\usepackage{amsmath}
\usepackage{amssymb}
\usepackage{bm}
\usepackage{}
\usepackage[
bookmarks=true,
colorlinks,
linkcolor=blue,
urlcolor=blue,
citecolor=blue,
plainpages=false,
pdfpagelabels,
final,
breaklinks=true
]{hyperref} % 参考文献
\usepackage{ulem}
\begin{document}
	
	\preprint{APS/123-QED}
	\title{Complete suppression of the non-dipole drift effect in high harmonic generation}
	\author{Hailang Wei$^{1}$}
\author{Xiaosong Zhu$^{1,3}$}\email{zhuxiaosong@hust.edu.cn}
%\author{Liang Li$^{1}$}
%\author{Wanzhu He$^{1}$}
%\author{Jie Long$^{1}$}
\author{Pengfei Lan$^{1,3}$}\email{pengfeilan@hust.edu.cn}
\author{Peixiang Lu$^{1,2,3}$}
% \author{Peixiang Lu$^{1,2,3}$}\email{lupeixiang@hust.edu.cn}

\affiliation{%
		$^1$ Wuhan National Laboratory for Optoelectronics and School of Physics,
		Huazhong University of Science and Technology, Wuhan 430074,
		China\\
		$^2$ Hubei Key Laboratory of Optical Information and  Pattern Recognition, Wuhan Institute of Technology, Wuhan 430205, China\\
            $^3$ Hubei Optical Fundamental Research Center,  Wuhan 430074, China     }

	%\date{\today}
	
	\begin{abstract}
 
     In high harmonic generation (HHG), non-dipole effects become increasingly significant at long driving wavelengths, as the magnetic field leads to a lateral drift of the continuum electron, which disrupts the electron recollision and inhibits the harmonic emission. To address this problem, we revisit the dynamics of the continuum electrons under electromagnetic fields in the HHG process and show that the magnetic effect on the drift includes a fundamental-frequency and a double-frequency component.  By adding an additional field to counteract the double-frequency effect caused by the magnetic field, we construct an effective linearly polarized field that recovers the recollision of all returning electrons to the parent ion. Consequently, the harmonic yield is restored and becomes the same as the result within the dipole approximation across the broad spectral range. This work provides a scheme that completely suppresses the non-dipole drift effect and fully compensates for the harmonic yield reduction, paving the way to efficiently generate coherent radiation in the range from extreme ultraviolet to soft x-ray and ultrashort pulses based on HHG.
     
     % electron dynamics along the laser propagation direction in an electromagnetic field can be approximated as a superposition of two components: (1) a fundamental-frequency component related to the ionization timing and (2) a double-frequency component.
     % we propose a scheme that completely suppresses the non-dipole drift upon recollision and thereby fully compensates for the harmonic yield reduction.
     % By adding an additional pulse to counteract the double-frequency effect caused by the magnetic field, we construct an effective linearly polarized field that recovers the recollision of all ionized electrons to the parent ion. The harmonic yield is restored and becomes identical to the result within the dipole approximation across the broad spectral range.
     % The effectiveness of this compensation scheme is confirmed by both classical and quantum simulations. 
     % This work provides a promising route to efficiently generate coherent radiation in the range from extreme ultraviolet to soft x-ray and ultrashort pulses based on HHG.
		
	\end{abstract}
	
	\maketitle
	
	\section{INTRODUCTION}

        High harmonic generation (HHG) is an extremely nonlinear optical phenomenon that arises from the interaction of intense laser fields with matter \cite{corkum_attosecond_2007,Attosecond2009,fleischer_spin_2014,long_polarization_2025}.
       % It provides a convenient and popular approach for generating extreme ultraviolet or x-ray attosecond pulses \cite{mairesse2003attosecond,popmintchev_bright_2012}.
        The abundant information in the high harmonic spectrum provides unique access to probe the structure and dynamics of the targets \cite{itatani2004tomographic,smirnova2009high,he2022filming,summers_realizing_2023,shirozhan_high-repetition-rate_2024,HeLixin2024}.
        Meanwhile, HHG serves as an excellent source of coherent extreme ultraviolet (XUV) radiation \cite{azoury_interferometric_2019,singh_intense_2021}. 
        By synthesizing broadband high-harmonic radiation, attosecond pulses can be produced in the time domain \cite{paul2001observation,hentschel2001attosecond,chini2014generation}.
        According to the cutoff law of HHG, the maximum harmonic frequency scales linearly with \( I\lambda^2 \), where \( I \) and \( \lambda \) are the intensity and wavelength of the driving laser, respectively \cite{Corkum1993,lewenstein_theory_1994}.
        To extend the cutoff of the high harmonics, using a driving field with a longer wavelength is a feasible method \cite{takahashi_attosecond_2013,li201753,gaumnitz2017streaking}.
        Long-wavelength drivers promote HHG to keV regime via exceptionally high harmonic orders, enabling ultrashort x-ray pulse production \cite{mairesse2003attosecond,popmintchev_bright_2012,HGC2013}.
        %More than ten years ago, it first was achieved to generate harmonics in the kev range with thousands of harmonic orders using a $3.9\ \mu\mathrm{m}$ driving field \cite{popmintchev_bright_2012}. 
        %The spectral range of HHG is efficiently extended by increasing the laser wavelength, without worrying about the ground state depletion by employing high intensity. 
        % The broader harmonic spectrum potentially supports to generate shorter pulses \cite{HGC2013,luo_ultra-short_2013} and reveals more information about the structure and the ultrafast dynamics of the target \cite{torres_revealing_2010,torres_extension_2010,summers_realizing_2023,shirozhan_high-repetition-rate_2024,HeLixin2024}.
        Numerical calculations show that pushing the driving wavelength further, for instance, to $9\ \mu\mathrm{m}$—could open the door to the generation of zeptosecond pulses \cite{HGC2013}.
        %further expanding the frontiers of ultrafast science.

      HHG can be understood through a three-step model involving ionization, acceleration, and recombination of the active electron \cite{Corkum1993,schafer_above_1993,lewenstein_theory_1994}. 
      Following ionization, the freed electron is accelerated by the oscillating electric field of the laser.
      While the magnetic field component induces a drift motion \cite{dammasch2001}, this effect is negligible for visible or near-infrared driving lasers at intensities around $10^{14}\ \mathrm{W/cm}^2$.
      %The drift caused by the magnetic field  \cite{dammasch2001} is negligible when using visible or near-infrared lasers at $10^{14}\ \mathrm{W/cm}^2$.
      However, at longer laser wavelengths and higher intensities, the effect of the magnetic field for the electron dynamics must be taken into account \cite{reiss_dipole-approximation_2000}.
      The Lorentz force induces significant electron drift \cite{verschl_relativistic_2007,jensen_nondipole_2020}, suppressing the recollision of the returning electron to the parent ion. This will lead to a significant reduction in the yield of high harmonics \cite{walser_high_2000,kylstra2001,ccc2002,zhu_non-dipole_2016}.

      Multiple schemes have been proposed to overcome this problem.
      Some methods employ exotic medium, such as antisymmetric molecular orbitals \cite{fischer_enhanced_2006}, positronium \cite{henrich_positronium_2004}, or exploit ultrahigh-intensity trains of attosecond pulses to drive plasma dynamics \cite{kohler_phase-matched_2011}.
      Other methods mitigate magnetic drift through regulating the waveform of the laser field, such as employing additional fundamental frequency field \cite{ccc2002}, quarter frequency field \cite{fischer_simulation_2004}, and applying two non-collinear circularly polarized beams \cite{pisanty_high_2018}.
      % For HHG in gas medium, several field waveform control schemes have been explored, such as the use of additional fundamental frequency pulses \cite{ccc2002}, quarter frequency pulses \cite{fischer_simulation_2004}, and the application of two non-collinear circularly polarized beams \cite{pisanty_high_2018}.
      However, these methods show limited controllability as they selectively modulate a subset of electron trajectories or merely confine electron motion to a finite range around the parent ion.
      Consequently, the reduced harmonic yield due to the non-dipole drift effect is partially compensated.
      % they only partially promote the HHG yield across a broad spectral range or recover the yield of high harmonics corresponding to specific frequency.
      These limitations underscore the urgent need to develop more effective schemes to suppress the non-dipole drift effect and compensate for the corresponding harmonic yield reduction.

      In our work, we revisit the dynamics of the continuum electrons under electromagnetic fields in the HHG process and propose a scheme that can completely suppress the non-dipole drift effect and recover the harmonic emission.
      We show that the electron acceleration along the laser propagation direction in the electromagnetic field can be approximated as a superposition of two components: (1) a fundamental-frequency component related to the ionization time and (2) a double-frequency component.
      By adding a double-frequency electric field to counteract the double-frequency component caused by the magnetic field, we create an effective linearly polarized field, thereby reviving recollision between the electron and the parent ion. The scheme is verified by numerical calculations based on both semiclassical and quantum models.
      % This method offers an effective route to generate bright attosecond and zeptosecond light sources using mid-infrared or even far-infrared lasers.
      
% \vspace{-3\baselineskip} % 向上压缩空白
   %The paper is structured as follows.
   %Section~\ref{theoretical_model} introduces the theoretical methods used in this study.
   %Section~\ref{RESULTS} introduces the compensation strategy and its operational framework. Classical and quantum models were used to verify the method's effectiveness.
   %Finally, we end this paper with a short summary and perspective in section~\ref{CONCLUSION}.
       % \newpage % 强制换页
        \section{THEORETICAL MODEL}\label{theoretical_model}

     %   \subsection{Classical equations of motion}\label{schematicalA}
    In this work, two theoretical frameworks are employed to characterize the electron-laser field interactions in the HHG process. Atomic units (a.u.) are used throughout this paper unless otherwise stated.
    
    The semiclassical model \cite{Corkum1993,schafer_above_1993} is used to study the electron trajectories in the electromagnetic fields of the driving laser during the HHG process.
    Following tunnel ionization, the continuum electrons behave as classical particles whose dynamics in the electromagnetic fields are governed by Newton's equations.
    By analyzing the trajectories of the electrons, the non-dipole drift can be intuitively understood. 
     To quantitatively characterize the HHG yield, this work additionally employs the non-dipole strong field approximation model to calculate the harmonic spectrum.
    % Within this framework, electrons are described in the form of electron wave packets, which transition from the ground state to the continuous state and move in the laser field, eventually returning to the ground state of the atom.
     In this framework, electron dynamics are described via laser-driven wave packet evolution.
     High-harmonic radiation is governed by coherent superposition of quantum paths, with dominant contributions from saddle-point trajectories \cite{Sali2001}.
     After using the saddle-point method, the time-dependent dipole moment is calculated by:
    \begin{equation}
    \mathbf{d}(t) \approx-2 \operatorname{Im} \sum_{t_{d}} a_{\mathbf{ion}}\left(t, t_{d}\right) a_{\mathbf{pr}}\left(t, t_{d}\right) \mathbf{a}_{\mathbf{rec}}^{*}\left(t, t_{d}\right)
   \end{equation}
   with the ionization, propagation, and recombination amplitudes given by \cite{ccc2002,zhu_non-dipole_2016}
    \begin{align}
    a_{\rm{ion}}\left(t, t_{d}\right) &= \frac{\left(8 I_{{p}}\right)^{5 / 4}}{8\left(2 s_{0} s_{2}\right)^{1 / 2}} \exp \left[-\frac{1}{3}\left(\frac{8 s_{0}^{3}}{s_{2}}\right)^{1 / 2}\right], \\
    a_{\rm{pr}}\left(t, t_{d}\right) &= \mathrm{C}\left({t}-t_{d}\right) \exp \left[-i S\left(p_{s}, t, t_{d}\right)\right], \\
    \mathbf{a}_{\rm{rec}}^{*}\left(t, t_{d}\right) &= \mathbf{d}_{\rm{rec}}^{*}\left[\pi\left(p_{s}, t\right)\right],
   \end{align}
  % \begin{equation}
 %   a_{\text {ion }}\left(t, t_{d}\right)=\frac{\left(8 I_{\mathrm{p}}\right)^{5 / 4}}{8\left(2 s_{0} s_{2}\right)^{1 / 2}} \exp \left[-\frac{1}{3}\left(\frac{8 s_{0}^{3}}{s_{2}}\right)^{1 / 2}\right],
  %\end{equation}
  %\begin{equation}
  %  a_{\mathrm{pr}}\left(t, t_{d}\right)=\mathrm{C}\left(\mathrm{t}-t_{d}\right) \exp \left[-i S\left(p_{s}, t, t_{d}\right)\right],
  %\end{equation}
  % \begin{equation}
  %  a_{r e c}^{*}\left(t, t_{d}\right)=d_{r e c}^{*}\left[\pi\left(p_{s}, t\right)\right],
  %\end{equation}
  where $s_{0} = I_{p} + \frac{1}{2} \pi_{k}^{2}(p_{s}, t_{d})$, $\mathrm{s}_{2} = E^{2}(\omega t_{d})$. $p_s$ and $t_d$ are the saddle momentum and approximate saddle time, respectively.
  \begin{align}
    C(\tau) &= (2 \pi)^{3 / 2}\left[(\xi + i \tau)^{3}\left[1 - \frac{1}{c^{2}}\left(\hat{\epsilon} \cdot p_{s}\right)^{2}\right]\right]^{-1 / 2}, \\
    S\left(\mathbf{p}, t, t^{\prime}\right) &= \frac{1}{2} \int_{t^{\prime}}^{t} dt^{\prime \prime} \left[\pi\left(\mathbf{p}, t^{\prime \prime}\right)\right]^{2} + I_{p}\left(t - t^{\prime}\right), \\
    \pi(\mathbf{p}, t) &= \mathbf{p} + \mathbf{A}(\omega t) + \frac{1}{c}\left[\mathbf{p} \cdot \mathbf{A}(\omega t) + \frac{1}{2} \mathbf{A}^{2}(\omega t)\right] \hat{\mathbf{k}}.
\end{align}
  %\begin{equation}
   % C(\tau)=(2 \pi)^{3 / 2}\left[(\xi+i \tau)^{3}\left[1-\frac{1}{c^{2}}\left(\hat{\epsilon} \cdot p_{s}\right)^{2}\right]\right]^{-1 / 2},
 % \end{equation}
 % \begin{equation}
   % S\left(\mathbf{p}, t, t^{\prime}\right)=\frac{1}{2} \int_{t^{\prime}}^{t} d t^{\prime \prime}\left[\pi\left(\mathbf{p}, t^{\prime \prime}\right)\right]^{2}+I_{p}\left(t-t^{\prime}\right),
 % \end{equation}
  %\begin{equation}
   % \pi(\mathbf{p}, t)=\mathbf{p}+\mathbf{A}(\omega t)+\frac{1}{c}\left[\mathbf{p} \cdot \mathbf{A}(\omega t)+\frac{1}{2} \mathbf{A}^{2}(\omega t)\right] \hat{\mathbf{k}}.
 % \end{equation}
    $\mathbf{A}(\omega t)$ is the vector potential of the laser. $\hat{\epsilon}$ represents the polarization vector of the driving laser that drives the HHG and $\hat{\mathbf{k}}$ represents the propagation direction of the field. $I_p$ is the ionization potential of the target atom.
    For hydrogen-like atoms, the transition dipole moment is given by
    \begin{equation}
    \mathbf{d}_{\rm{rec}}(\mathbf{q})=i \frac{2^{7 / 2}}{\pi}\left(2 I_{p}\right)^{5 / 4} \frac{\mathbf{q}}{\left(\mathbf{q}^{2}+2 I_{p}\right)^{3}}.
    \end{equation}
    Here, $\mathbf{q}=\pi(\mathbf{p}, t)$.
    This study involves electric fields in the $x$ and $z$ directions.
    %  Considering the electric fields in the $x$ and $z$ directions,
    The solution of the saddle-point equation $\nabla_{\mathbf{p}} S(\mathbf{p}, t, t^{\prime}) = \rm{0}$ gives the following $x$ and $z$ components of the saddle momentum:
    \begin{align}
    p_{\mathrm{sx}} &= -\frac{\alpha_{x}^{[1]}}{\tau}, \\
    p_{\mathrm{sz}} &= -\frac{\alpha_{z}^{[1]} + \frac{1}{c}\left(-\frac{1}{\tau}\right)\left[\alpha_{x}^{[1]}\right]^{2} + \frac{3}{2c} \alpha_{z}^{[2]} + \frac{1}{2c} \alpha_{x}^{[2]}}{\tau + \frac{2}{c} \alpha_{z}^{[1]}},
    \end{align}
    %+ \frac{1}{c^{2}} \alpha_{z}^{[2]}
    %\begin{equation}
   % p_{st \mathrm{x}}=-\frac{\alpha_{x}^{[1]}}{\tau},
  %  \end{equation}
   % \begin{equation}
    %p_{stz}=-\frac{\alpha_{z}^{[1]}+\frac{1}{c}\left(-\frac{1}{\tau}\right)\left[\alpha_{x}^{[1]}\right]^{2}+\frac{3}{2 c} \alpha_{z}^{[2]}+\frac{1}{2 c} \alpha_{x}^{[2]}}{\tau+\frac{2}{c} \alpha_{z}^{[1]}+\frac{1}{c^{2}} \alpha_{z}^{[2]}},
   % \end{equation}
    where $\tau=t-t^{\prime}$, $\alpha_{i}^{[\mathrm{n}]}=\int_{t^{\prime}}^{t} A_{i}^{n} d t^{\prime \prime}$. The expression for $p_{\mathrm{sz}}$ neglects the higher order terms in $1/c$.
    % The saddle-point approximation along the $z$-axis excludes the non-dominant terms.

      Finally, the harmonic spectrum is obtained from the Fourier transform of the dipole acceleration $\ddot{\mathbf{d}}(t)$
    \begin{equation}
    \mathbf{E}_{\rm{XUV}}(\Omega)=\int \ddot{\mathbf{d}}(t) \exp (-i \Omega t) d t,
    \end{equation}
    \begin{equation}
    S_{\rm{I}}(\Omega)=\left|\mathbf{E}_{\rm{XUV}}(\Omega)\right|^{2}.
    \end{equation}

    The results in the dipole approximation can be obtained by setting 1/c = 0.

     \section{RESULTS AND DISCUSSION}\label{RESULTS}

     \subsection{Analysis for the complete non-dipole drift effect suppression}\label{schematicalIIIA}

     In this section, we employ the semiclassical model to analyze the trajectories of electrons in electromagnetic fields and propose a method that can fully suppress the non-dipole drift in the HHG process.
     % The classical trajectory approach offers clear physical insight into the drift-compensation mechanism.
     % In this part, we propose a scheme for completely compensating magnetic drift by constructing an equivalent linear polarization field.

      The driving laser, which drives HHG, is polarized along the $x$ direction and propagates along the $z$ direction, with its electric field expressed as:
      \begin{equation}
      E_{x} = E_{0} \cos(\omega t).
      \end{equation}
     The magnetic field is $B_y(t)=E_x(t)/c$. 
     According to Newton's equations, classical electron trajectories satisfy:
    \begin{align}
    \ddot{x} &= - \big( E_x - \dot{z} B_y \big) \label{eq:x}, \\
    \ddot{z} &= - \dot{x} B_y \label{eq:z}.
    \end{align}
   % \begin{equation}\label{co}
   % \ddot{\mathbf{r}} = - ( \mathbf{E} + \dot{\mathbf{r}} \times \mathbf{B} ).
   % \end{equation} 
    The solution of these equations describes the time-dependent trajectory of the electron ionized at time $t_i$ from the origin:
\begin{align}
x(t) &= \frac{E_0}{\omega^2} \left[ \cos(\omega t) - \cos(\omega t_i) \right] + \frac{E_0}{\omega} \sin(\omega t_i) (t - t_i) \label{eq:xt}, \\
z(t) &= -\frac{E_{0}^{2}}{8 \omega^{3} c}\left[\sin (2 \omega t)-\sin \left(2 \omega t_{i}\right)\right] \nonumber \\
&\quad +\frac{E_{0}^{2}}{\omega^{3} c} \sin \left(\omega t_{i}\right)\left[\cos (\omega t)-\cos \left(\omega t_{i}\right)\right] \nonumber \\
&\quad +\left[\frac{E_{0}^{2}}{4 \omega^{2} c} \cos \left(2 \omega t_{i}\right)+\frac{E_{0}^{2}}{\omega^{2} c} \sin ^{2}\left(\omega t_{i}\right)\right]\left(t-t_{i}\right) .\label{eq:zt}
\end{align}
    When solving Eq.~(\ref{eq:x}), the $\dot{z} B_y$ contribution to acceleration is neglected, as it is negligible compared to $E_x$.
   % The electron undergoes tunnel ionization at time $t_i$ (starting from the origin) and moves under the laser field. 
   While tunnel ionized at $t_i$ and accelerated in the continuum, the recombining time $t_r$ of the electron is judged by the second occurrence of $x(t_r)=0$.
   Namely, at the recombining time $t_r$, the electron drifts $z{(t_r)}$ with the parent ion.
   Excessive electron drift induced by the Lorentz force disrupts the recollision process and reduces the HHG yield.
     To recover the HHG yield, conventional methods resort to minimizing the drift $z(t)$. However, since the electron motion in the $z$-direction is quite complicated, as shown by Eq.~(\ref{eq:zt}), it is challenging to completely eliminate the drift $z(t)$.

     We note that, elimination of the drift $z(t)$ in the $z$-direction is not a necessary condition to completely restore the recollision. The recollision can be guaranteed as long as the ionized electron is driven in a straight line, ensuring that it always heads directly toward the parent ion upon return. 
     % More importantly, this is unnecessary. In our view, we believe that the key issue lies in the simultaneous return of electrons in both the $x$ and $z$ directions.
% We will establish a scheme for complete compensation of the HHG yield reduction due to the nondipole drift following this perspective.  
We will establish a scheme for complete suppression of the non-dipole drift effect following this perspective. 

    To better analyze the drift in the $z$ direction, we recast Eq.~(\ref{eq:z}) as:
    \begin{align} \label{eq:z_motion}
    \ddot{z} &= -\left[
              \frac{E_0^2}{\omega c}\sin(\omega t_i)\cos(\omega t)- \frac{E_0^2}{2\omega c}\sin(2\omega t) \right] \nonumber \\
    \end{align}
   Eq.~(\ref{eq:z_motion}) indicates that the acceleration of electrons along the $z$ direction under the effect of the magnetic field can be divided into two parts: (1) a fundamental-frequency component related to the ionization time and (2) a double-frequency component.
   % Comparing Eq.~(\ref{eq:x}) (ignoring the magnetic field-related terms $\dot{z} B_y$), if the double frequency component is counteracted, the $x$ and $z$ directions will achieve motion of the same frequency and phase, that is, an effective linear polarization field will be constructed.
    Comparing Eq.~(\ref{eq:x}) (dropping the negligible $\dot{z} B_y$ term), if the double-frequency component is counteracted, the $x$ and $z$ components of the electron motion share the same frequency and phase. Namely, the electron moves in a straight line as driven by an effective linear polarization field in the $x-z$ plane.

    For this purpose, we consider a weak control field propagating in the $x$ direction and linearly polarized in the $z$ direction. The electric field takes the form:
    \begin{equation}
      E_{z} = E_{z0} \cos(n \omega t + \varphi)
    \end{equation}
    When this field meets the following conditions:
    \begin{align} 
    E_{z0} &= \frac{E_0^2}{2\omega c} \label{eq:Ex0} \\
    n      &= 2 \label{eq:n} \\
    \varphi &= -\frac{\pi}{2} \label{eq:phi}
    \end{align}
    The equation of motion of the electrons along the $z$ direction becomes:
\begin{equation} \label{eq:z2_motion}
\begin{split}
\ddot{z} &= -\frac{E_0}{\omega c}\sin(\omega t_i)E_x \\
         &= -\gamma(t_i)E_x
\end{split}
\end{equation}
    Thus, the trajectory of the electron ionized at time $t_i$ satisfies:
    \begin{equation} \label{eq:z_relation}
    z(t, t_i) = \gamma(t_i) x(t, t_i).
    \end{equation}
    % Where the coefficient $\gamma$ is equal to $\frac{E_0}{\omega c}$.
    Note that $\gamma(t_i)$ is a constant related to $t_i$. The result means that the electrons move in an effective linearly polarized field, whose polarization direction varies according to the ionization time of individual electrons.
    % Electrons can return to the parent ion simultaneously along the $x$ and $z$ directions to undergo the recollision process. 
    Thus, the non-dipole drift effect is completely suppressed, and the corresponding HHG yield reduction can be fully compensated.
      \subsection{Numerical results for the non-dipole drift suppression}\label{schematicalIIA}  

      %In this section, we present the results under the classic model.
     To verify the above discussions, we numerically calculate the trajectories of electrons in the electromagnetic fields.
     We consider electrons ionized from Ar ($I_p=0.58$ a.u.) with zero initial momenta. The driving field is linearly polarized in the $x$ direction and propagates in the $z$ direction, with a peak intensity of $3 \times 10^{14}\ \mathrm{W/cm}^2$ and wavelength of $4\ \mu\mathrm{m}$. 
     The control electric field is linearly polarized in the $z$ direction and propagates in the $x$ direction. It has a peak intensity of $2.63 \times 10^{11}~\mathrm{W}/\mathrm{cm}^2$, a wavelength of $2~\mu\mathrm{m}$, and a phase of $-\pi/2$, satisfying all the requirements specified in Eqs.~(\ref{eq:Ex0}), (\ref{eq:n}), and (\ref{eq:phi}).

    \begin{figure}[ht]
		\centerline{
		\includegraphics[scale=0.41]{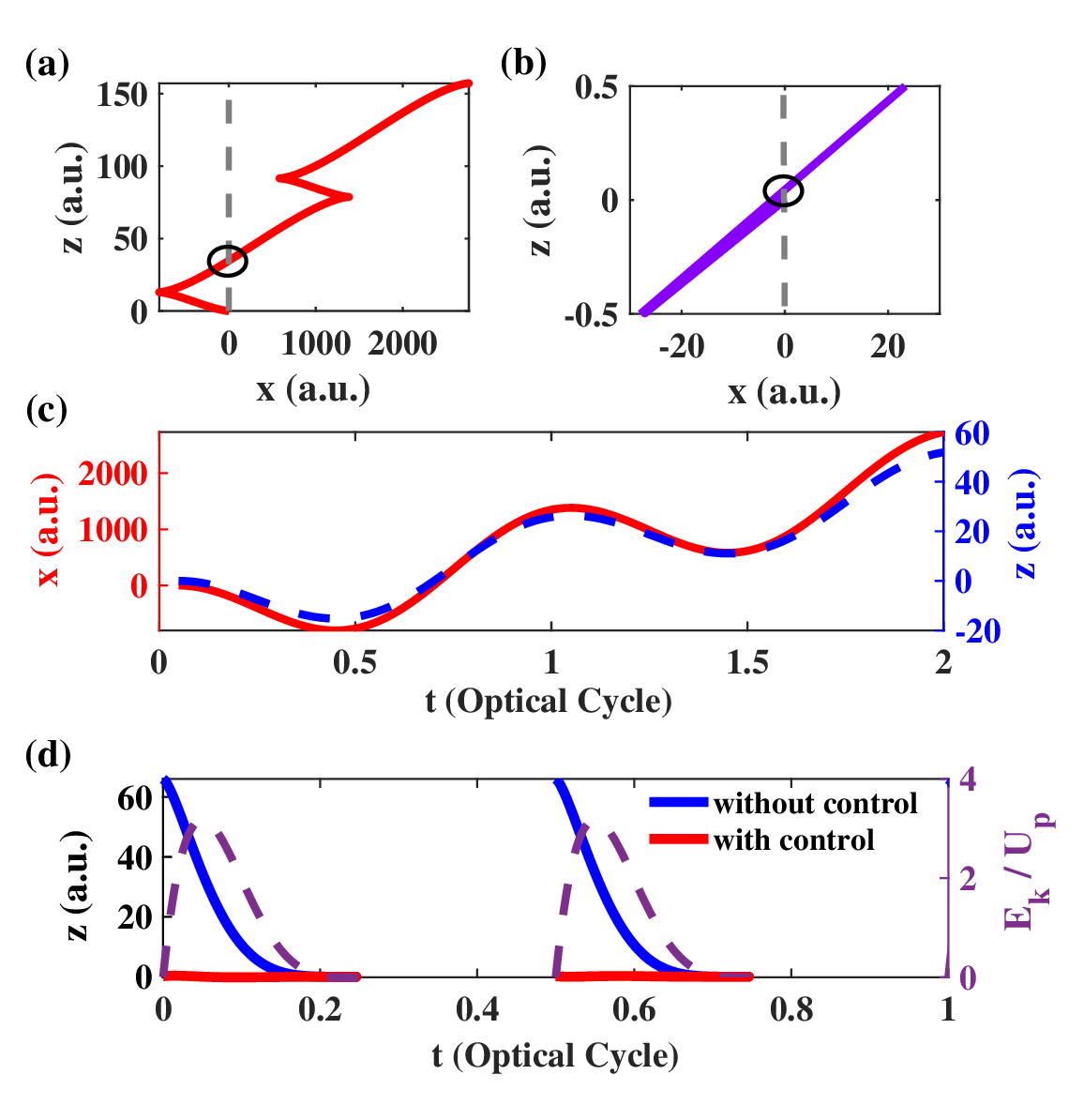}}
		\caption{\label{bandstructure1}
       (a) Electron trajectory with return kinetic energy 3.17$U_p$ in the presence of only the driving field. 
       (b) Electron trajectory under the combined driving and control fields, with the same return kinetic energy (3.17$U_p$).
       The black circle marks that the electron drifts along the $z$ when it returns in the $x$ direction for the first time.
       (c) % Time-dependent electron motion along the $x$ direction (red curve) and $z$ direction (blue curve) separately after adding the control pulse.
       $x$ and $z$ components of the time-dependent electron motion after adding the control field.
       (d) The drift amplitude (left axis) along the $z$ direction with (blue curve) and without (red curve) the control field for electrons with ionization time varying over the optical cycle.
       The purple dotted line traces the kinetic energy of return electrons  (right axis) as a function of ionization time.}
	\end{figure} 
     
     Figures~\hyperref[bandstructure1]{\textcolor{blue}{\ref*{bandstructure1}(a)}} and \hyperref[bandstructure1]{\textcolor{blue}{\ref*{bandstructure1}(b)}} show the trajectories of electrons returning with kinetic energy of $3.17 U_p$ (maximum kinetic energy), with only the driving field and with both the driving and control fields, respectively.
     $U_p = E_0^2 / (4 \omega^2)$ refers to the ponderomotive energy of electrons in the laser field.
     When only the driving field, is present, the electrons drift $35~\text{a.u.}$ along the $z$ direction.
     %upon recollision with the parent ion.
     With a large drift, the recollision between electrons and the parent ion is prevented, and the harmonic emissions are diminished.
     
\begin{figure}[ht]
		\centerline{
		\includegraphics[scale=0.39]{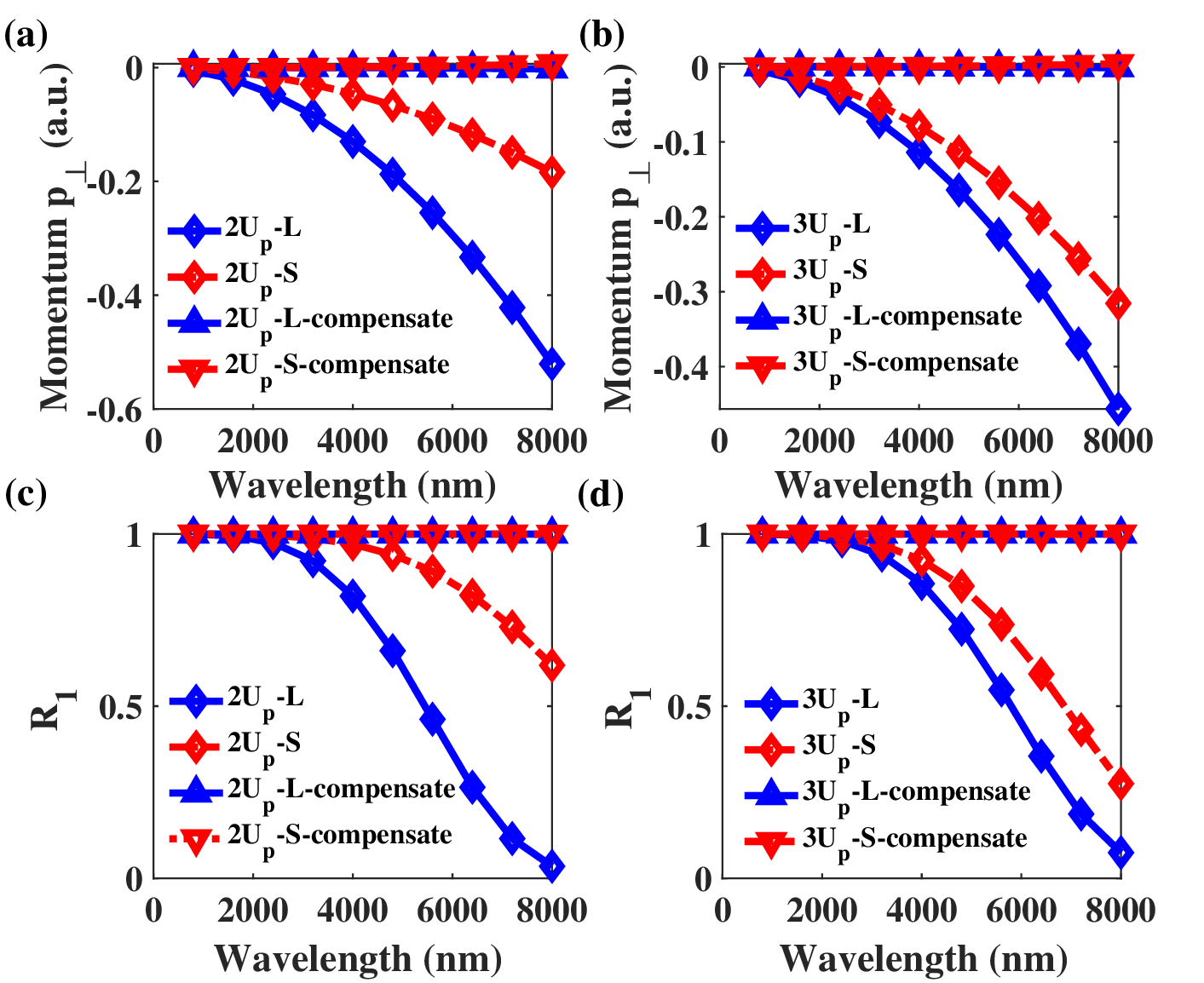}}
		\caption{\label{bandstructure2}
        (a-b) Initial lateral momentum needed for recollision for electrons with return kinetic energies of (a) $2U_p$ and (b) $3U_p$ as a function of the driving laser wavelength.
        (c-d) The ratio $R_1$ for electrons with return kinetic energies of (c) 2$U_p$ and (d) 3$U_p$.
        Diamond markers: driving field alone; triangular markers: driving field plus control field.}
	\end{figure}

     After adding the control field, the two-dimensional trajectory of the electron lies in nearly a straight line, as shown in Fig.~\hyperref[bandstructure1]{\textcolor{blue}{\ref*{bandstructure1}(b)}}. %,  implying that the electron is moving in an effective linearly polarized field.
     When electrons return to the parent ion, the drift along the $z$ direction is less than $1~\text{a.u.}$. 
     This forms a sharp contrast with the electron trajectory shown in Fig.~\hyperref[bandstructure1]{\textcolor{blue}{\ref*{bandstructure1}(a)}}.
    Figure~\hyperref[bandstructure2]{\textcolor{blue}{\ref*{bandstructure1}(c)}} shows the $x$ and $z$ components of the time-dependent electron motion after adding the control field.
    Clearly, the electron motion in the $x$ and $z$ directions shares the same frequency and phase. This is in good agreement with what Eq.~(\ref{eq:z_relation}) expected.
    To verify that our scheme suppresses the non-dipole drift for electrons with all return kinetic energies, we calculate the drift amplitude with and without the control field for electrons with ionization time varying over the optical cycle ($T = 2\pi/\omega_0$).
    The results are shown in Fig.~\hyperref[bandstructure1]{\textcolor{blue}{\ref*{bandstructure1}(d)}}.
    After adding the control field (red curve), for electrons ionized at any time, the drift along the propagation is less than $1~\text{a.u.}$ upon recollision, while the uncompensated case (blue curve) exhibits significant lateral drift.
    The complete suppression of the drift ensures the recollsion of the electrons to the parent ion without deviation, which is expected to completely compensate for the HHG yield reduction due to the non-dipole effect.%scale=0.45 width=\columnwidth

% The ratio $R_1$ of ionization rate non-dipole to the dipole approximation contributed by the long-orbit and short-orbit electrons with return kinetic energies of (e) 2$U_p$ and (f) 3$U_p$.

     The effectiveness of our suppression scheme can be evaluated within the semiclassical theoretical framework by analyzing the lateral momentum $p_{\perp}$ needed for a recollision.
    When a nonzero lateral initial momentum $p_{\perp}$ of the ionized electron is taken into account, the displacement obtained by the electron due to the initial momentum and the drift caused by the magnetic field effect cancel each other out during the acceleration step, thus allowing the electron to undergo a recollision process with the parent ion.
     However, these electrons with nonzero initial momentum have a lower ionization rate, which will eventually lead to a decrease in harmonic yield.
     Specifically, after the electrons undergo tunneling ionization, the weight of each trajectory contributing to HHG is \cite{wang_momentum_2018,li_elliptically_2022}  
    \begin{equation}
    W\left(t_{i}, p_{\perp}\right)=w_{t}\left(t_{i}\right) w_{p}\left(p_{\perp}\right).
    \end{equation}
    $w_t$ is the ionization rate for electrons with zero initial momentum at $t_i$ given by the Ammosov-Delone-Krainov model \cite{ammosov_tunnel_1986}.
    $w_p$ is the dependence of the ionization rate on the initial momentum $p_{\perp}$ \cite{delone_energy_1991}, which is given by:
    %The decrease in harmonic yield can be described by the following formula:
    \begin{equation}
    w_{p}=\exp \left[-\frac{2\left(2 I_{p}+p_{\perp}^{2}\right)^{3 / 2}}{3\left|E\left(t_{i}\right)\right|}\right].
    \end{equation}
    Compared with electrons of $p_\perp=0$, electrons with non-zero $p_\perp$ correspond to lower $w_p$. The contribution of these electron trajectories to HHG decreases exponentially.

        \begin{figure*}[ht]
		\centerline{
		\includegraphics[scale=0.5]{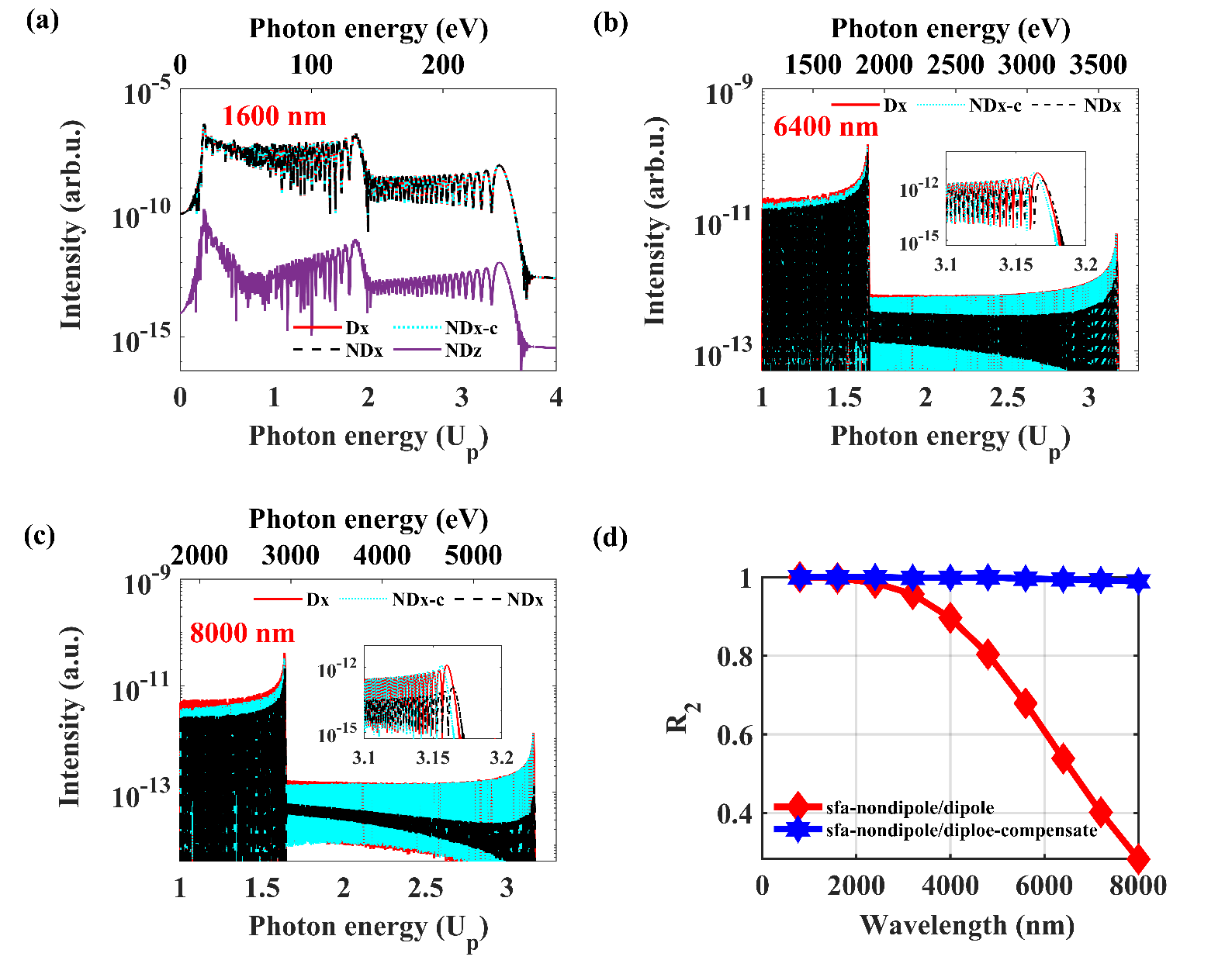}}
		\caption{\label{bandstructure3}
		(a-c) Harmonic spectra for $I = 3 \times 10^{14}\ \mathrm{W/cm}^2$ and $\lambda$ = 1600, 6400, $8000\,\text{nm}$, respectively. Dx: $x$ component of harmonics with only the driving field under dipole approximation; NDx (NDz): $x$ ($z$) components of non-dipole harmonics with only the driving field; NDx-c: $x$ component of non-dipole harmonics with both the driving field and control field.
        (d) Ratio $R_2$ for which the non-dipole harmonics are calculated with (blue curve) and without (red curve) adding the control field, respectively, at different wavelengths.}
        % (d) Ratio $R_2$ of harmonic yields from non-dipole to dipole approximation with (blue curve) and without (red curve) adding the control field at different wavelengths.}
	\end{figure*}

     Figures~\hyperref[bandstructure2]{\textcolor{blue}{\ref*{bandstructure2}(a)}}  and \hyperref[bandstructure1]{\textcolor{blue}{\ref*{bandstructure2}(b)}} respectively show the initial lateral momentum required to collide with the parent ion (origin) for electrons with return kinetic energies of $2U_p$ and $3U_p$  as a function of the driving laser wavelength. Both the long and short trajectories are considered, respectively.
    % Results calculated with only the driving pulse are marked by rhombuses, while those that include the control pulse are indicated by triangles.
     The results show that, when only the driving field is applied (diamond markers), in order to return to the parent ion, the initial lateral momentum required increases significantly with the increase of wavelength.
     However, after adding the control field (triangular markers), electrons only need a very small lateral momentum to recollide with the parent ion, which is shown in the figure as the horizontal curves at nearly zero across all wavelengths.
     Furthermore, we define the ratio $R_1$ to evaluate the suppression effect:
    % \vspace{-0.45em} % 调整负值以减少间距
\begin{equation}
    R_1=\frac{w_{p}\left(p_{\perp}, t_{i}\right)}{w_{p}\left(0, t_{i}\right)}.
\end{equation}
    In Figs.~\hyperref[bandstructure2]{\textcolor{blue}{\ref*{bandstructure2}(c)}} and \hyperref[bandstructure2]{\textcolor{blue}{\ref*{bandstructure2}(d)}}, we calculate the ratio $R_1$ for electrons with return kinetic energies of 2$U_p$ and 3$U_p$, respectively. 
    In the absence of the control field, the ratio decreases dramatically with the wavelength, suggesting a significant reduction of the HHG yield.
    %Meanwhile, we also found that the ratio contributed by long-orbit electrons and short-orbit high-energy recovery electrons decreased particularly significantly.
    Upon introduction of the control field, the ratio approaches unity, indicating that our scheme can completely suppress the effect of the non-dipole drift irrespective of the driving laser wavelength.
    % This is attributed to the effective linearly polarized electric field we constructed, which results in almost no lateral drift when electrons return to the parent ion.

        \subsection{Harmonic yield compensation}\label{schematicalIIB}

    To quantitatively characterize the compensation effect for the HHG yield, we employ the non-dipole strong-field approximation model to calculate the harmonic spectrum. For comparison, we also calculate the harmonic spectrum under the dipole approximation with only the driving field, establishing the baseline for assessing the compensation efficacy.

     %  Through comparative analysis of the harmonic yields between nondipole and dipole approximations prior to the auxiliary field implementation, we established the wavelength-dependent characteristics of harmonic generation. Furthermore, by comparing the harmonic yields of nondipoles and dipoles after adding the auxiliary electric field, we have further validated the effectiveness of our proposed scheme in compensating for the harmonic yield reduction induced by magnetic field effects.

   % The calculations were performed for argon, with $I_p=0.58$ a.u.. 
    The driving field is polarized along the $x$ direction and propagates along the $z$ direction. We consider a $\sin^2$ laser field with the full width of three optical cycles ($3T_0$). The electric field is expressed as:
    \begin{equation}
    \mathbf{E}_x(t)=E_{0} \sin ^{2}\left(\frac{\pi t}{3 T_{0}}\right) \cos (\omega t) \hat{\mathbf{x}}.
    \end{equation}
    The peak intensity of the driving electric field is $3 \times 10^{14}\ \mathrm{W/cm}^2$.
    The control field polarized in the $z$ direction and propagating in the $x$ direction has the form:
     \begin{equation}
    \mathbf{E}_z(t)=\frac{E_{0}^{2}}{2 \omega c} \sin ^{2}\left(\frac{\pi t}{3 T_{0}}\right) \cos \left(2 \omega t-\frac{\pi}{2}\right) \hat{\mathbf{k}}.
    \end{equation}
    Its parameters satisfy Eqs.~\eqref{eq:Ex0}, \eqref{eq:n}, and \eqref{eq:phi}.
    The driving laser wavelength $\lambda$ is scanned from $800\,\text{nm}$ to $8000\,\text{nm}$ in steps of $\Delta\lambda = 800\,\text{nm}$.
    Representative results for three specific wavelengths ($1600\,\text{nm}$, $6400\,\text{nm}$, $8000\,\text{nm}$) are shown in Figure.~\ref{bandstructure3}.
    Since the $z$ component harmonics are $10^3$--$10^5$ times weaker than the $x$ component (see Fig.~\hyperref[bandstructure3]{\textcolor{blue}{\ref*{bandstructure3}(a)}}) and the $z$ harmonic component cannot be detected in a typical HHG experiment, this component is not shown in Figs.~\hyperref[bandstructure3]{\textcolor{blue}{\ref*{bandstructure3}(b-c)}}.  % The harmonic yield detected in the experiment is determined by the $x$ component.

    Compared with the results of dipole approximation, when only the driving field is present, the harmonic yield taking account of the nondipole effects decreases significantly with increasing wavelength, especially in the high-frequency region.
    In addition, the harmonics exhibit a blueshift as shown in the insets of Figs.~\hyperref[bandstructure3]{\textcolor{blue}{\ref*{bandstructure3}(b-c)}}, which has been found in previous research works \cite{zhu_non-dipole_2016,walser_high_2000}.
  After adding the control field, the harmonics yield shows a significant enhancement over the wide spectral range, as shown by the cyan curves in Figs.~\hyperref[bandstructure3]{\textcolor{blue}{\ref*{bandstructure3}(b-c)}}.
  % The reduction in harmonic yield caused by the non-dipole drift within the harmonic order range under study has been completely compensated.
    % After adding the control pulse, the harmonics are shown by the blue dotted lines in Fig.~\hyperref[bandstructure3]{\textcolor{blue}{\ref*{bandstructure3}(b-c)}}. The harmonics yield shows a significant enhancement over a wide photon energy range.

    To quantify the compensation for harmonic yield, we define the ratio $R_2$:
\begin{equation}
R_2=\frac{\int_{q_\mathrm{min}}^{q_\mathrm{max}} d q I_{x}^{N D}}{\int_{q_\mathrm{min}}^{q_\mathrm{max}} d q I_{x}^{D}}.
\end{equation}
    $I_{x}^{ND}$ denotes the harmonic intensity taking account of the non-dipole effects, calculated with: (i) driving field alone (black curves in Figs.~\hyperref[bandstructure3]{\textcolor{blue}{\ref*{bandstructure3}(a-c)}}) and (ii) driving field and control field (cyan curves).
    $I_{x}^{D}$ represents the harmonic intensity obtained with only the driving field under the dipole approximation (red curves).
    The integration covers orders $q_\mathrm{min}$ to $q_\mathrm{max}$ corresponding to photon energies of 2.2$U_p$ and 3$U_p$ , respectively.
    Figure~\hyperref[bandstructure3]{\textcolor{blue}{\ref*{bandstructure3}(d)}} presents the calculated ratio $R_2$ as a function of the driving laser wavelength.
    When only the driving field is applied (red diamonds), because of the non-dipole effect, the ratio decreases monotonically with increasing wavelength.
    Upon addition of the control field to the driving field, the ratio $R_2$ reaches unity.
    %across the investigated spectral range.
    % The reduction in harmonic yield caused by the non-dipole effects within the harmonic order range under study has been completely compensated.
    In particular, this compensation scheme is applicable over a broad range of driving laser wavelengths.
    % this correction scheme exhibits a broad applicability range for the driving laser wavelength.

\begin{figure}[ht]
		\centerline{
		\includegraphics[scale=0.43]{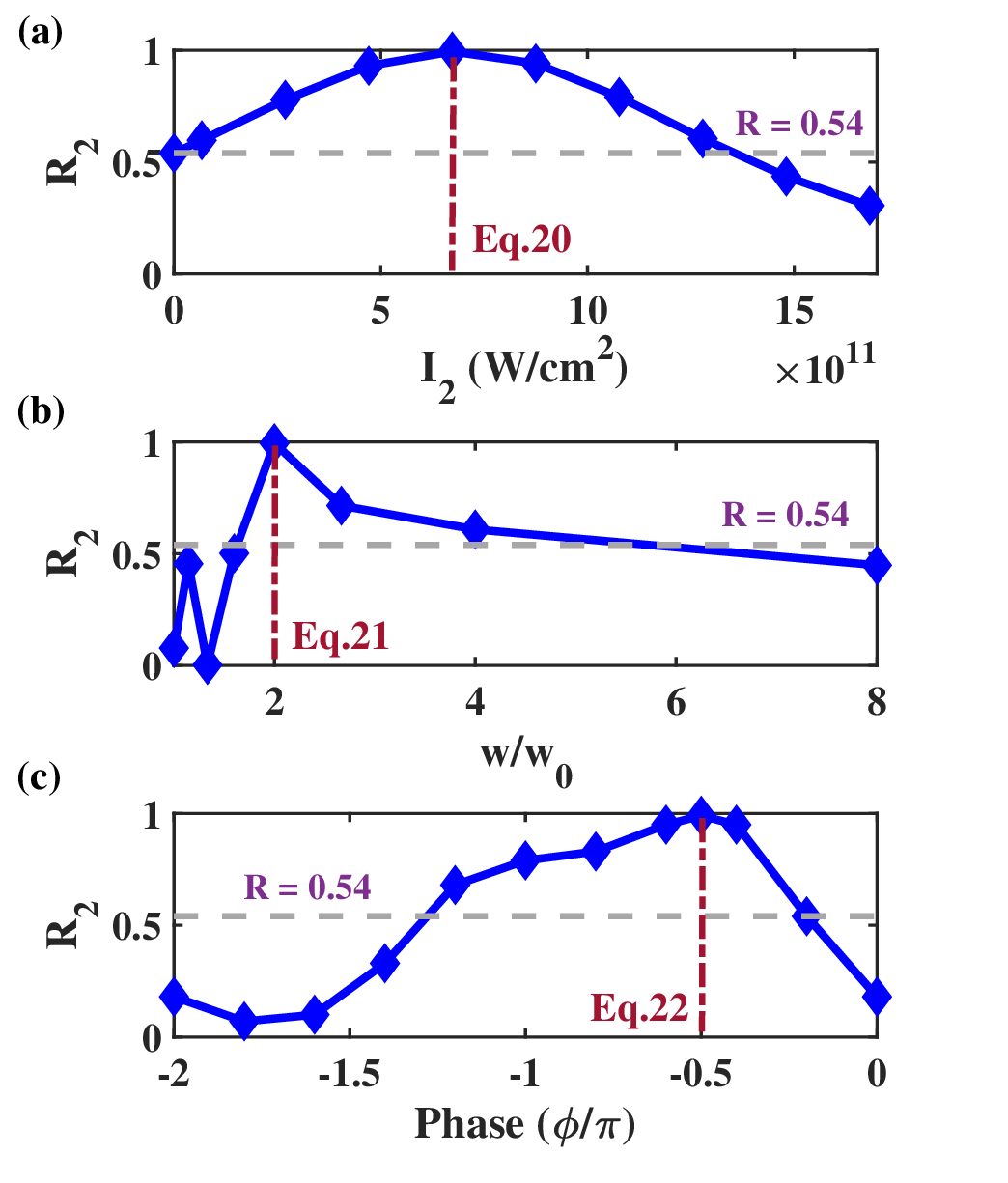}}
		\caption{\label{bandstructure4} 
     Ratio $R_2$ for varying control field parameters (a) peak intensity, (b) frequency, and (c) phase, respectively.}
    %Vary the control field parameters (a) peak intensity, (b) frequency, (c) phase and calculate the ratio for each condition. The wavelength of the driving field is fixed at $6400\,\text{nm}$ and the peak intensity is $3 \times 10^{14}\ \mathrm{W/cm}^2$. Grey dotted line $R_2=0.54$ in the figure represents the ratio without adding the control field under non-dipole.}
    \end{figure}

   To further validate our theory discussed in Sec~\ref{schematicalIIIA}, we vary the control field parameters and calculate the ratio for each condition in Figs.~\hyperref[bandstructure4]{\textcolor{blue}{\ref*{bandstructure4}(a-c)}}. Specifically, we fix the wavelength of the driving field at 6400 nm, and the peak intensity at $3 \times 10^{14}\ \mathrm{W/cm}^2$. 
   For the control field, we respectively scan one of the parameters (peak intensity, wavelength, and phase) while keeping the other two constant according to Eqs.~(\ref{eq:Ex0}), (\ref{eq:n}), and (\ref{eq:phi}).
   % Eqs.\~(\textbackslash{}ref\{eq:Ex0\}), (\textbackslash{}ref\{eq:n\}), and (\textbackslash{}ref\{eq:phi\}). 
   %and scan the peak intensity, wavelength, and phase of the control pulse, respectively. The rest two unscanned parameters take the values according to Eqs.~(\ref{eq:Ex0}), (\ref{eq:n}) and (\ref{eq:phi}).
   % In Fig.~\hyperref[bandstructure4]{\textcolor{blue}{\ref*{bandstructure4}(a)}}, the control pulse intensity is varied (Eq.~\eqref{eq:Ex0}) while its frequency and phase are fixed according to Eqs.~(\eqref{eq:n}) and (\eqref{eq:phi}), respectively. Similarly, Fig.~\hyperref[bandstructure4]{\textcolor{blue}{\ref*{bandstructure4}(b)}} explores the frequency dependence at fixed intensity and phase satisfying Eqs.~(\ref{eq:Ex0}) and (\ref{eq:phi}), whereas Fig.~\hyperref[bandstructure4]{\textcolor{blue}{\ref*{bandstructure4}(c)}} examines the phase variation with intensity and frequency held constant according to Eqs.~(\ref{eq:Ex0}) and (\ref{eq:n}).
   The gray dotted line ($R_2$= 0.54) corresponds to the result without control field.
   Figure~\hyperref[bandstructure4]{\textcolor{blue}{\ref*{bandstructure4}(a)}} shows that the compensation effect reaches its peak, where $R_2$ reaches unity, at the control field intensity of $6.73 \times 10^{11}~\mathrm{W/cm}^2$, in agreement with the prediction of Eq.~(\ref{eq:Ex0}).
   % Figure~\hyperref[bandstructure4]{\textcolor{blue}{\ref*{bandstructure4}(a)}} shows that the compensation effect peaks \textcolor{red}{when $R_2$ reaches unity,} at the control laser intensity of $6.73 \times 10^{11}~\mathrm{W/cm}^2$, consistent with the prediction by Eq.~(\ref{eq:Ex0}).
   Figures~\hyperref[bandstructure4]{\textcolor{blue}{\ref*{bandstructure4}(b)}} and \hyperref[bandstructure4]{\textcolor{blue}{\ref*{bandstructure4}(c)}} show that the compensation effect peaks ($R_2$ reaches unity) only when the frequency and phase of the control field satisfy Eqs.~(\ref{eq:n}) and (\ref{eq:phi}), respectively. The coincidence between the analytical discussions and the numerical results strongly underpins the theoretical basis of our proposed scheme.
   % Importantly, these peak-related laser parameters comply with all the standards established by the scheme we proposed (section \ref{schematicalIIIA}), further supporting our theory.
  % This is all thanks to the equivalent linearly polarized field we constructed, which almost completely eliminates the lateral drift effect caused by the magnetic field. When electrons return, they collide head-on with the parent ion, restarting the emission of harmonics.

   \section{CONCLUSION}\label{CONCLUSION}

   In conclusion, we analyze the interaction between the continuum electrons and electromagnetic fields in the HHG process and propose a scheme to completely suppress the non-dipole drift effect. Our analysis shows that the non-dipole effect consists of a fundamental-frequency component and a double-frequency component. By adding the control field to counteract the double frequency component, we construct an effective linear polarization field, enabling all returning electrons to head directly to the parent ion without lateral drift.  As a result, the reduction of HHG yield due to the non-dipole drift can be fully compensated across the broad spectral range.
   The effectiveness of this compensation scheme is confirmed based on both classical and quantum simulations.
   This work provides a promising route to enhance the efficiency of HHG with long driving laser wavelengths, enabling more efficient generation of coherent XUV and soft x-ray radiation and ultrashort pulses.
 
        \section*{ACKNOWLEDGMENTS}
      This work was supported by National Key Research and Development Program (Grant No. 2023YFA1406800) and the National Natural Science Foundation of China (NSFC) (Grants No. 12174134, No. 12225406, and No. 12021004). The computation is completed in the HPC Platform of Huazhong University of Science and Technology.

	%\bibliography{weihailang}

\begin{thebibliography}{99}


\bibitem{corkum_attosecond_2007}
P. B. Corkum and F. Krausz, Nat. Phys. 3, 381 (2007).
% P.~B. Corkum and F.~Krausz, ``Attosecond science,'' \emph{Nat. Phys.}, vol.~3, no.~6, pp. 381--387, Jun. 2007. [Online]. Available: \url{https://www.nature.com/articles/nphys620}

\bibitem{Attosecond2009}
F. Krausz and M. Ivanov, Rev. Mod. Phys. 81, 163 (2009).
% F.~Krausz and M.~Ivanov, ``Attosecond physics,'' \emph{Rev. Mod. Phys.}, vol.~81, pp. 163--234, Feb 2009. [Online]. Available: \url{https://link.aps.org/doi/10.1103/RevModPhys.81.163}

\bibitem{fleischer_spin_2014}
A. Fleischer, O. Kfir, T. Diskin, P. Sidorenko, and O. Cohen, Nat. Photonics 8, 543 (2014).
% A.~Fleischer, O.~Kfir, T.~Diskin, P.~Sidorenko, and O.~Cohen, ``Spin angular momentum and tunable polarization in high-harmonic generation,'' \emph{Nat. Photonics}, vol.~8, no.~7, pp. 543--549, Jul. 2014. [Online]. Available: \url{https://www.nature.com/articles/nphoton.2014.108}

\bibitem{long_polarization_2025}
J. Long, X. Zhu, C. Zhai, W. Li, W. He, L. He, P. Lan, and P. Lu, Ultrafast Sci. 5, 0079 (2025).
% J.~Long, X.~Zhu, C.~Zhai, W.~Li, W.~He, L.~He, P.~Lan, and P.~Lu, ``Polarization {Control} in {High} {Harmonic} {Generation} {Using} {Molecular} {Structures} in {Nonaligned} {Molecules},'' \emph{Ultrafast Sci.}, vol.~5, p. 0079, Jan. 2025. [Online]. Available: \url{https://spj.science.org/doi/10.34133/ultrafastscience.0079}

\bibitem{itatani2004tomographic}
J.~Itatani, J.~Levesque, D.~Zeidler, H.~Niikura, H.~P{\'e}pin, J.-C. Kieffer, P.~B. Corkum, and D.~M. Villeneuve, Nature 432, 867 (2004).
% J.~Itatani, J.~Levesque, D.~Zeidler, H.~Niikura, H.~P{\'e}pin, J.-C. Kieffer, P.~B. Corkum, and D.~M. Villeneuve, ``Tomographic imaging of molecular orbitals,'' \emph{Nature}, vol. 432, no. 7019, pp. 867--871, 2004.

\bibitem{smirnova2009high}
 O. Smirnova, Y. Mairesse, S. Patchkovskii, N. Dudovich,
 D. Villeneuve, P. Corkum, and M. Y. Ivanov, Nature 460,
 972 (2009).
% O.~Smirnova, Y.~Mairesse, S.~Patchkovskii, N.~Dudovich, D.~Villeneuve, P.~Corkum, and M.~Y. Ivanov, ``High harmonic interferometry of multi-electron dynamics in molecules,'' \emph{Nature}, vol. 460, no. 7258, pp. 972--977, 2009.

\bibitem{he2022filming}
L. He, S. Sun, P. Lan, Y. He, B. Wang, P. Wang, X. Zhu, L. Li, W. Cao, P. Lu, and C. D. Lin, Nat. Commun. 13, 4595 (2022).
% L.~He, S.~Sun, P.~Lan, Y.~He, B.~Wang, P.~Wang, X.~Zhu, L.~Li, W.~Cao, P.~Lu, and C.~D. Lin, ``Filming movies of attosecond charge migration in single molecules with high harmonic spectroscopy,'' \emph{Nat. Commun.}, vol.~13, p. 4595, Aug 2022.

\bibitem{summers_realizing_2023}
A.~M. Summers, S.~Severino, M.~Reduzzi, T.~P.~H. Sidiropoulos, D.~E. Rivas, N.~Di~Palo, H.-W. Sun, Y.-H. Chien, I.~León, B.~Buades, S.~L. Cousin, S.~M. Teichmann, T.~Mey, K.~Mann, B.~Keitel, E.~Plönjes, D.~K. Efetov, H.~Schwoerer, and J.~Biegert, Ultrafast Sci. 3, 0004 (2023).
% A.~M. Summers, S.~Severino, M.~Reduzzi, T.~P.~H. Sidiropoulos, D.~E. Rivas, N.~Di~Palo, H.-W. Sun, Y.-H. Chien, I.~León, B.~Buades, S.~L. Cousin, S.~M. Teichmann, T.~Mey, K.~Mann, B.~Keitel, E.~Plönjes, D.~K. Efetov, H.~Schwoerer, and J.~Biegert, ``Realizing {Attosecond} {Core}-{Level} {X}-ray {Spectroscopy} for the {Investigation} of {Condensed} {Matter} {Systems},'' \emph{Ultrafast Sci.}, vol.~3, p. 0004, Jan. 2023. [Online]. Available: \url{https://spj.science.org/doi/10.34133/ultrafastscience.0004}

\bibitem{shirozhan_high-repetition-rate_2024}
M.~Shirozhan, S.~Mondal, T.~Grósz, B.~Nagyillés, B.~Farkas, A.~Nayak, N.~Ahmed, I.~Dey, S.~C. De~Marco, K.~Nelissen, M.~Kiss, L.~G. Oldal, T.~Csizmadia, Z.~Filus, M.~De~Marco, S.~Madas, M.~U. Kahaly, D.~Charalambidis, P.~Tzallas, E.~Appi, R.~Weissenbilder, P.~Eng-Johnsson, A.~L’Huillier, Z.~Diveki, B.~Major, K.~Varjú, and S.~Kahaly, Ultrafast Sci. 4, 0067 (2024).
% M.~Shirozhan, S.~Mondal, T.~Grósz, B.~Nagyillés, B.~Farkas, A.~Nayak, N.~Ahmed, I.~Dey, S.~C. De~Marco, K.~Nelissen, M.~Kiss, L.~G. Oldal, T.~Csizmadia, Z.~Filus, M.~De~Marco, S.~Madas, M.~U. Kahaly, D.~Charalambidis, P.~Tzallas, E.~Appi, R.~Weissenbilder, P.~Eng-Johnsson, A.~L’Huillier, Z.~Diveki, B.~Major, K.~Varjú, and S.~Kahaly, ``High-{Repetition}-{Rate} {Attosecond} {Extreme} {Ultraviolet} {Beamlines} at {ELI} {ALPS} for {Studying} {Ultrafast} {Phenomena},'' \emph{Ultrafast Sci.}, vol.~4, p. 0067, Jan. 2024. [Online]. Available: \url{https://spj.science.org/doi/10.34133/ultrafastscience.0067}

\bibitem{HeLixin2024}
L. He, C. H. Yuen, Y. He, S. Sun, E. Goetz, A.-T. Le,
 Y. Deng, C. Xu, P. Lan, P. Lu, and C. D. Lin, Phys. Rev.
 Lett. 133, 023201 (2024).
% L.~He, C.~H. Yuen, Y.~He, S.~Sun, E.~Goetz, A.-T. Le, Y.~Deng, C.~Xu, P.~Lan, P.~Lu, and C.~D. Lin, ``Ultrafast picometer-resolved molecular structure imaging by laser-induced high-order harmonics,'' \emph{Phys. Rev. Lett.}, vol. 133, p. 023201, Jul 2024. [Online]. Available: \url{https://link.aps.org/doi/10.1103/PhysRevLett.133.023201}


\bibitem{azoury_interferometric_2019}
D.~Azoury, O.~Kneller, M.~Krüger, B.~D. Bruner, O.~Cohen, Y.~Mairesse, and N.~Dudovich, Nat. Photonics 13,
 198 (2019).
% D.~Azoury, O.~Kneller, M.~Krüger, B.~D. Bruner, O.~Cohen, Y.~Mairesse, and N.~Dudovich, ``Interferometric attosecond lock-in measurement of extreme-ultraviolet circular dichroism,'' \emph{Nat. Photonics}, vol.~13, no.~3, pp. 198--204, Mar. 2019. [Online]. Available: \url{https://www.nature.com/articles/s41566-019-0350-5}

\bibitem{singh_intense_2021}
M.~Singh, M.~A. Fareed, V.~Strelkov, A.~N. Grum-Grzhimailo, A.~Magunov, A.~Laramée, F.~Légaré, and T.~Ozaki, Optica 8, 1122 (2021).
% M.~Singh, M.~A. Fareed, V.~Strelkov, A.~N. Grum-Grzhimailo, A.~Magunov, A.~Laramée, F.~Légaré, and T.~Ozaki, ``Intense quasi-monochromatic resonant harmonic generation in the multiphoton ionization regime,'' \emph{Optica}, vol.~8, no.~8, p. 1122, Aug. 2021. [Online]. Available: \url{https://opg.optica.org/abstract.cfm?URI=optica-8-8-1122}

\bibitem{paul2001observation}
P.-M. Paul, E.~S. Toma, P.~Breger, G.~Mullot, F.~Aug{\'e}, P.~Balcou, H.~G. Muller, and P.~Agostini, Science 292,
 1689 (2001).
% P.-M. Paul, E.~S. Toma, P.~Breger, G.~Mullot, F.~Aug{\'e}, P.~Balcou, H.~G. Muller, and P.~Agostini, ``Observation of a train of attosecond pulses from high harmonic generation,'' \emph{Science}, vol. 292, no. 5522, pp. 1689--1692, 2001.

\bibitem{hentschel2001attosecond}
M.~Hentschel, R.~Kienberger, C.~Spielmann, G.~A. Reider, N.~Milosevic, T.~Brabec, P.~Corkum, U.~Heinzmann, M.~Drescher, and F.~Krausz, Nature 414, 509 (2001).
% M.~Hentschel, R.~Kienberger, C.~Spielmann, G.~A. Reider, N.~Milosevic, T.~Brabec, P.~Corkum, U.~Heinzmann, M.~Drescher, and F.~Krausz, ``Attosecond metrology,'' \emph{Nature}, vol. 414, no. 6863, pp. 509--513, 2001.

\bibitem{chini2014generation}
M. Chini, K. Zhao, and Z. Chang, Nat. Photonics 8, 178
 (2014).
% M.~Chini, K.~Zhao, and Z.~Chang, ``The generation, characterization and applications of broadband isolated attosecond pulses,'' \emph{Nat. Photonics}, vol.~8, no.~3, pp. 178--186, 2014.


\bibitem{Corkum1993}
P. B. Corkum, Phys. Rev. Lett. 71, 1994 (1993).
% P.~B. Corkum, ``Plasma perspective on strong-field multiphoton ionization,'' \emph{Phys. Rev. Lett.}, vol.~71, no.~13, pp. 1994--1997, Sep. 1993.

\bibitem{lewenstein_theory_1994}
 M. Lewenstein, P. Balcou, M. Y. Ivanov, A. L’Huillier,
 and P. B. Corkum, Phys. Rev. A 49, 2117 (1994).
% M.~Lewenstein, P.~Balcou, M.~Y. Ivanov, A.~L’Huillier, and P.~B. Corkum, ``Theory of high-harmonic generation by low-frequency laser fields,'' \emph{Phys. Rev. A}, vol.~49, no.~3, pp. 2117--2132, Mar. 1994. [Online]. Available: \url{https://link.aps.org/doi/10.1103/PhysRevA.49.2117}


\bibitem{takahashi_attosecond_2013}
 E.~J. Takahashi, P.~Lan, O.~D. Mücke, Y.~Nabekawa, and K.~Midorikawa, Nat. Commun. 4, 2691 (2013).
% E.~J. Takahashi, P.~Lan, O.~D. Mücke, Y.~Nabekawa, and K.~Midorikawa, ``Attosecond nonlinear optics using gigawatt-scale isolated attosecond pulses,'' \emph{Nat. Commun.}, vol.~4, no.~1, p. 2691, Oct. 2013. [Online]. Available: \url{https://www.nature.com/articles/ncomms3691}

\bibitem{li201753}
 J.~Li, X.~Ren, Y.~Yin, K.~Zhao, A.~Chew, Y.~Cheng, E.~Cunningham, Y.~Wang, S.~Hu, Y.~Wu, M.~Chini, and Z.~Chang, Nat. Commun. 8, 186 (2017).
% J.~Li, X.~Ren, Y.~Yin, K.~Zhao, A.~Chew, Y.~Cheng, E.~Cunningham, Y.~Wang, S.~Hu, Y.~Wu, M.~Chini, and Z.~Chang, ``53-attosecond x-ray pulses reach the carbon k-edge,'' \emph{Nat. Commun.}, vol.~8, no.~1, p. 186, 2017.

\bibitem{gaumnitz2017streaking}
 T.~Gaumnitz, A.~Jain, Y.~Pertot, M.~Huppert, I.~Jordan, F.~Ardana-Lamas, and H.~J. W{\"o}rner, Opt. Express 25, 27506 (2017).
% T.~Gaumnitz, A.~Jain, Y.~Pertot, M.~Huppert, I.~Jordan, F.~Ardana-Lamas, and H.~J. W{\"o}rner, ``Streaking of 43-attosecond soft-x-ray pulses generated by a passively cep-stable mid-infrared driver,'' \emph{Opt. Express}, vol.~25, no.~22, pp. 27\,506--27\,518, 2017.


\bibitem{mairesse2003attosecond}
Y.~Mairesse, A.~de~Bohan, L.~J. Frasinski, H.~Merdji, L.~C. Dinu, P.~Monchicourt, P.~Breger, M.~Kovacev, R.~Taieb, B.~Carre, H.~G. Muller, P.~Agostini, and P.~Sali{\`e}res, Science 302, 1540 (2003).
% Y.~Mairesse, A.~de~Bohan, L.~J. Frasinski, H.~Merdji, L.~C. Dinu, P.~Monchicourt, P.~Breger, M.~Kovacev, R.~Taieb, B.~Carre, H.~G. Muller, P.~Agostini, and P.~Sali{\`e}res, ``Attosecond synchronization of high-harmonic soft x-rays,'' \emph{Science}, vol. 302, no. 5650, pp. 1540--1543, Nov 2003.

\bibitem{popmintchev_bright_2012}
T.~Popmintchev, M.-C. Chen, D.~Popmintchev, P.~Arpin, S.~Brown, S.~Ališauskas, G.~Andriukaitis, T.~Balčiunas, O.~D. Mücke, A.~Pugzlys, A.~Baltuška, B.~Shim, S.~E. Schrauth, A.~Gaeta, C.~Hernández-García, L.~Plaja, A.~Becker, A.~Jaron-Becker, M.~M. Murnane, and H.~C. Kapteyn, Science 336, 1287 (2012).
% T.~Popmintchev, M.-C. Chen, D.~Popmintchev, P.~Arpin, S.~Brown, S.~Ališauskas, G.~Andriukaitis, T.~Balčiunas, O.~D. Mücke, A.~Pugzlys, A.~Baltuška, B.~Shim, S.~E. Schrauth, A.~Gaeta, C.~Hernández-García, L.~Plaja, A.~Becker, A.~Jaron-Becker, M.~M. Murnane, and H.~C. Kapteyn, ``Bright {Coherent} {Ultrahigh} {Harmonics} in the {keV} {X}-ray {Regime} from {Mid}-{Infrared} {Femtosecond} {Lasers},'' \emph{Science}, vol. 336, no. 6086, pp. 1287--1291, Jun. 2012. [Online]. Available: \url{https://www.science.org/doi/10.1126/science.1218497}

\bibitem{HGC2013}
C.~Hern\'andez-Garc\'{\i}a, J.~A. P\'erez-Hern\'andez, T.~Popmintchev, M.~M. Murnane, H.~C. Kapteyn, A.~Jaron-Becker, A.~Becker, and L.~Plaja, Phys. Rev. Lett. 111, 033002 (2013).
% C.~Hern\'andez-Garc\'{\i}a, J.~A. P\'erez-Hern\'andez, T.~Popmintchev, M.~M. Murnane, H.~C. Kapteyn, A.~Jaron-Becker, A.~Becker, and L.~Plaja, ``Zeptosecond high harmonic kev x-ray waveforms driven by mid-infrared laser pulses,'' \emph{Phys. Rev. Lett.}, vol. 111, no.~3, p. 033002, Jul. 2013.

\bibitem{schafer_above_1993}
K.~J. Schafer, B.~Yang, L.~F. DiMauro, and K.~C. Kulander, Phys. Rev. Lett. 70, 1599 (1993).
% K.~J. Schafer, B.~Yang, L.~F. DiMauro, and K.~C. Kulander, ``Above threshold ionization beyond the high harmonic cutoff,'' \emph{Phys. Rev. Lett.}, vol.~70, no.~11, pp. 1599--1602, Mar. 1993. [Online]. Available: \url{https://link.aps.org/doi/10.1103/PhysRevLett.70.1599}

\bibitem{dammasch2001}
M.~Dammasch, M.~Dörr, U.~Eichmann, E.~Lenz, and W.~Sandner, Phys. Rev. A 64, 061402 (2001).
% M.~Dammasch, M.~Dörr, U.~Eichmann, E.~Lenz, and W.~Sandner, ``Relativistic laser-field-drift suppression of nonsequential multiple ionization,'' \emph{Phys. Rev. A}, vol.~64, no.~6, p. 061402, Nov. 2001. [Online]. Available: \url{https://link.aps.org/doi/10.1103/PhysRevA.64.061402}

\bibitem{reiss_dipole-approximation_2000}
H. R. Reiss, Phys. Rev. A 63, 013409 (2000).
% H.~R. Reiss, ``Dipole-approximation magnetic fields in strong laser beams,'' \emph{Phys. Rev. A}, vol.~63, no.~1, p. 013409, Dec. 2000. [Online]. Available: \url{https://link.aps.org/doi/10.1103/PhysRevA.63.013409}

\bibitem{verschl_relativistic_2007}
M. Verschl and C. H. Keitel, Phys. Rev. ST Accel. Beams
 10, 024001 (2007).
% M.~Verschl and C.~H. Keitel, ``\BIBforeignlanguage{en}{Relativistic classical and quantum dynamics in intense crossed laser beams of various polarizations},'' \emph{\BIBforeignlanguage{en}{Phys. Rev. ST Accel. Beams}}, vol.~10, no.~2, p. 024001, Feb. 2007. [Online]. Available: \url{https://link.aps.org/doi/10.1103/PhysRevSTAB.10.024001}

\bibitem{jensen_nondipole_2020}
S.~V.~B. Jensen, M.~M. Lund, and L.~B. Madsen, Phys.
 Rev. A 101, 043408 (2020).
% S.~V.~B. Jensen, M.~M. Lund, and L.~B. Madsen, ``Nondipole strong-field-approximation {Hamiltonian},'' \emph{Phys. Rev. A}, vol. 101, no.~4, p. 043408, Apr. 2020. [Online]. Available: \url{https://link.aps.org/doi/10.1103/PhysRevA.101.043408}


\bibitem{walser_high_2000}
M.~W. Walser, C.~H. Keitel, A.~Scrinzi, and T.~Brabec,
 Phys. Rev. Lett. 85, 5082 (2000).
% M.~W. Walser, C.~H. Keitel, A.~Scrinzi, and T.~Brabec, ``High {Harmonic} {Generation} {Beyond} the {Electric} {Dipole} {Approximation},'' \emph{Phys. Rev. Lett.}, vol.~85, no.~24, pp. 5082--5085, Dec. 2000. [Online]. Available: \url{https://link.aps.org/doi/10.1103/PhysRevLett.85.5082}


\bibitem{kylstra2001}
N.~J. Kylstra, R.~M. Potvliege, and C.~J. Joachain, J.
Phys. B 34, L55 (2001).
% N.~J. Kylstra, R.~M. Potvliege, and C.~J. Joachain, ``Photon emission by ions interacting with short intense laser pulses: beyond the dipole approximation,'' \emph{J. Phys. B}, vol.~34, no.~3, p. L55, Feb. 2001. [Online]. Available: \url{https://dx.doi.org/10.1088/0953-4075/34/3/101}

\bibitem{ccc2002}
 C.~C. Chirilă, N.~J. Kylstra, R.~M. Potvliege, and C.~J. Joachain, Phys. Rev. A 66, 063411 (2002).
% C.~C. Chirilă, N.~J. Kylstra, R.~M. Potvliege, and C.~J. Joachain, ``Nondipole effects in photon emission by laser-driven ions,'' \emph{Phys. Rev. A}, vol.~66, no.~6, p. 063411, Dec. 2002. [Online]. Available: \url{https://link.aps.org/doi/10.1103/PhysRevA.66.063411}

\bibitem{zhu_non-dipole_2016}
X. Zhu and Z. Wang, Opt. Commun. 365, 125 (2016).
% X.~Zhu and Z.~Wang, ``Non-dipole effects on high-order harmonic generation towards the long wavelength region,'' \emph{Opt. Commun.}, vol. 365, pp. 125--132, Apr. 2016. [Online]. Available: \url{https://linkinghub.elsevier.com/retrieve/pii/S0030401815303291}



\bibitem{fischer_enhanced_2006}
 R.~Fischer, M.~Lein, and C.~H. Keitel, Phys. Rev. Lett.
 97, 143901 (2006).
% R.~Fischer, M.~Lein, and C.~H. Keitel, ``Enhanced {Recollisions} for {Antisymmetric} {Molecular} {Orbitals} in {Intense} {Laser} {Fields},'' \emph{Phys. Rev. Lett.}, vol.~97, no.~14, p. 143901, Oct. 2006. [Online]. Available: \url{https://link.aps.org/doi/10.1103/PhysRevLett.97.143901}


\bibitem{henrich_positronium_2004}
B.~Henrich, K.~Z. Hatsagortsyan, and C.~H. Keitel, Phys.
 Rev. Lett. 93, 013601 (2004).
% B.~Henrich, K.~Z. Hatsagortsyan, and C.~H. Keitel, ``Positronium in {Intense} {Laser} {Fields},'' \emph{Phys. Rev. Lett.}, vol.~93, no.~1, p. 013601, Jul. 2004. [Online]. Available: \url{https://link.aps.org/doi/10.1103/PhysRevLett.93.013601}


\bibitem{kohler_phase-matched_2011}
M.~C. Kohler, M.~Klaiber, K.~Z. Hatsagortsyan, and C.~H. Keitel, Europhys. Lett. 94, 14002 (2011).
% M.~C. Kohler, M.~Klaiber, K.~Z. Hatsagortsyan, and C.~H. Keitel, ``Phase-matched coherent hard {X}-rays from relativistic high-order harmonic generation,'' \emph{Europhys. Lett.}, vol.~94, no.~1, p. 14002, Apr. 2011. [Online]. Available: \url{https://dx.doi.org/10.1209/0295-5075/94/14002}


\bibitem{fischer_simulation_2004}
R.~Fischer, A.~Staudt, and C.~Keitel, Comput. Phys.
 Commun. 157, 139 (2004).
% R.~Fischer, A.~Staudt, and C.~Keitel, ``Simulation of atomic quantum dynamics in combined intense laser and weak electric fields,'' \emph{Comput. Phys. Commun.}, vol. 157, no.~2, pp. 139--146, Feb. 2004. [Online]. Available: \url{https://linkinghub.elsevier.com/retrieve/pii/S0010465503005174}

\bibitem{pisanty_high_2018}
E.~Pisanty, D.~D. Hickstein, B.~R. Galloway, C.~G. Durfee, H.~C. Kapteyn, M.~M. Murnane, and M.~Ivanov, New J. Phys. 20, 053036 (2018).
% E.~Pisanty, D.~D. Hickstein, B.~R. Galloway, C.~G. Durfee, H.~C. Kapteyn, M.~M. Murnane, and M.~Ivanov, ``High harmonic interferometry of the {Lorentz} force in strong mid-infrared laser fields,'' \emph{New J. Phys.}, vol.~20, no.~5, p. 053036, May 2018. [Online]. Available: \url{https://iopscience.iop.org/article/10.1088/1367-2630/aabb4d}

\bibitem{Sali2001}
P.~Salières, B.~Carré, L.~L. Déroff, F.~Grasbon, G.~G. Paulus, H.~Walther, R.~Kopold, W.~Becker, D.~B. Milošević, A.~Sanpera, and M.~Lewenstein, Science 292,
 902 (2001).
% P.~Salières, B.~Carré, L.~L. Déroff, F.~Grasbon, G.~G. Paulus, H.~Walther, R.~Kopold, W.~Becker, D.~B. Milošević, A.~Sanpera, and M.~Lewenstein, ``Feynman's path-integral approach for intense-laser-atom interactions,'' \emph{Science}, vol. 292, no. 5518, pp. 902--905, May 2001.


\bibitem{wang_momentum_2018}
D. Wang, X. Zhu, L. Li, X. Zhang, X. Liu, P. Lan, and
 P. Lu, Phys. Rev. A 98, 053410 (2018).
% D.~Wang, X.~Zhu, L.~Li, X.~Zhang, X.~Liu, P.~Lan, and P.~Lu, ``Momentum gate for tunneling electrons with a circularly polarized control field,'' \emph{Phys. Rev. A}, vol.~98, no.~5, p. 053410, Nov. 2018. [Online]. Available: \url{https://link.aps.org/doi/10.1103/PhysRevA.98.053410}

\bibitem{li_elliptically_2022}
 W. Li, X. Zhu, P. Lan, and P. Lu, Phys. Rev. A 106,
 043115 (2022).
% W.~Li, X.~Zhu, P.~Lan, and P.~Lu, ``Elliptically polarized attosecond pulse generation by corotating bicircular laser fields,'' \emph{Phys. Rev. A}, vol. 106, no.~4, p. 043115, Oct. 2022. [Online]. Available: \url{https://link.aps.org/doi/10.1103/PhysRevA.106.043115}


\bibitem{ammosov_tunnel_1986}
M. V. Ammosov, N. B. Delone, and V. P. Krainov, SPIE
 0664, 138 (1986).
% M.~V. Ammosov, N.~B. Delone, and V.~P. Krainov, ``Tunnel ionization of complex atoms and atomic ions in electromagnetic field,'' \emph{SPIE}, vol. 0664, no.~64, pp. 138--141, Oct. 1986.

\bibitem{delone_energy_1991}
 N. B. Delone and V. P. Krainov, J. Opt. Soc. Am. B 8,
 1207 (1991)
% N.~B. Delone and V.~P. Krainov, ``Energy and angular electron spectra for the tunnel ionization of atoms by strong low-frequency radiation,'' \emph{J. Opt. Soc. Am. B}, vol.~8, no.~6, pp. 1207--1211, Jun. 1991. [Online]. Available: \url{https://opg.optica.org/josab/abstract.cfm?uri=josab-8-6-1207}


\end{thebibliography}

\end{document}